\documentclass[conference]{IEEEtran}
\IEEEoverridecommandlockouts
\usepackage{cite}
\usepackage{amsmath,amssymb,amsfonts}
\usepackage{algorithmic}
\usepackage{graphicx}
\usepackage{textcomp}
\usepackage{xcolor}
\usepackage{eucal}
\usepackage{textcomp}
\usepackage{multicol}
\usepackage{multirow}
\usepackage{url}
\usepackage{booktabs}
\usepackage{subfigure}

\def\BibTeX{{\rm B\kern-.05em{\sc i\kern-.025em b}\kern-.08em
    T\kern-.1667em\lower.7ex\hbox{E}\kern-.125emX}}
\begin{document}

\newcommand{\tech}{\textsc{MATT}}
\newcommand{\knn}{\textbf{KNN}}
\newcommand{\lr}{\textbf{LR}}
\newcommand{\svm}{\textbf{SVM}}
\newcommand{\mlp}{\textbf{MLP}}
\newcommand{\fcn}{\textbf{fcn}}
\newcommand{\crnn}{\textbf{CRNN}}
\newcommand{\cgrnn}{\textbf{CRNN-TF}}
\newcommand{\chroma}{\textbf{Chroma}}
\newcommand{\tonnetc}{\textbf{Tonnetc}}
\newcommand{\mfcc}{\textbf{MFCC}}
\newcommand{\centroid}{\textbf{Spec. Centroid}}
\newcommand{\bandwidth}{\textbf{Spec. Bandwidth}}
\newcommand{\contrast}{\textbf{Spec. Contrast}}
\newcommand{\rolloff}{\textbf{Spec. Rolloff}}
\newcommand{\rms}{\textbf{RMS Energy}}
\newcommand{\zcr}{\textbf{Zero-crossing Rate}}
\pdfoutput=1

\title{MATT: A Multiple-instance Attention Mechanism for Long-tail Music Genre Classification
}
\author{\IEEEauthorblockN{Xiaokai Liu$^*$\thanks{$^*$ Corresponding Author}, Menghua Zhang$^\S$}\\
	\IEEEauthorblockA{$^*$School of Cyber Science and Engineering\\
		$^{\S}$  School of Energy and Power Engineering
		\\ Huazhong University of Science and Technology, Wuhan, China\\
		\texttt{\{liuxk, iemhzhang\}@hust.edu.cn}}
}
\maketitle

\begin{abstract}


Imbalanced music genre classification is a crucial task in the Music Information Retrieval (MIR) field for identifying the long-tail, data-poor genre based on the related music audio segments, which is very prevalent in real-world scenarios. Most of the existing models are designed for class-balanced music datasets, resulting in poor performance in accuracy and generalization when identifying the music genres at the tail of the distribution. Inspired by the success of introducing Multi-instance Learning (MIL) in various classification tasks, we propose a novel mechanism named Multi-instance Attention (\tech)\footnote{
\ Github Link: https://github.com/JohannesLiu/Music-Genre-Classification} to boost the performance for identifying tail classes. Specifically, we first construct the bag-level datasets by generating the album-artist pair bags. Second, we leverage neural networks to encode the music audio segments. Finally, under the guidance of a multi-instance attention mechanism, the neural network-based models could select the most informative genre to match the given music segment. Comprehensive experimental results on a large-scale music genre benchmark dataset with long-tail distribution demonstrate \tech\ significantly outperforms other state-of-the-art baselines.

\end{abstract}


\section{Introduction}


Music Genre Classification (MGC)~\cite{DBLP:journals/eswa/CorreaR16}, which aims to identify the genres of given music segments, is an essential task in the research field of Music Information Retrieval (MIR)

Recently, inspired by the progress achieved by Deep Learning (DL) techniques, researchers ~\cite{DBLP:conf/cbmi/SenacPMP17,DBLP:conf/nips/LeePLN09} have widely applied DL techs in the fields of MIR and achieved promising results in various tasks, especially MGC. However, since the neural networks are mostly data-hungry, their performance is heavily subject to the scale and quality of training data. Although such techs achieve very good results on common genres, their performances degrade drastically while extracting long-tail genres, which indicates they routinely suffer from data insufficiency. Moreover, most of the previous works only focus on the benchmarks with class-balanced data. These facts lead the classification of long-tail music genres to be a very challenging problem.

\begin{figure}[htbp]
	\centerline{\includegraphics[width=1.0\columnwidth]{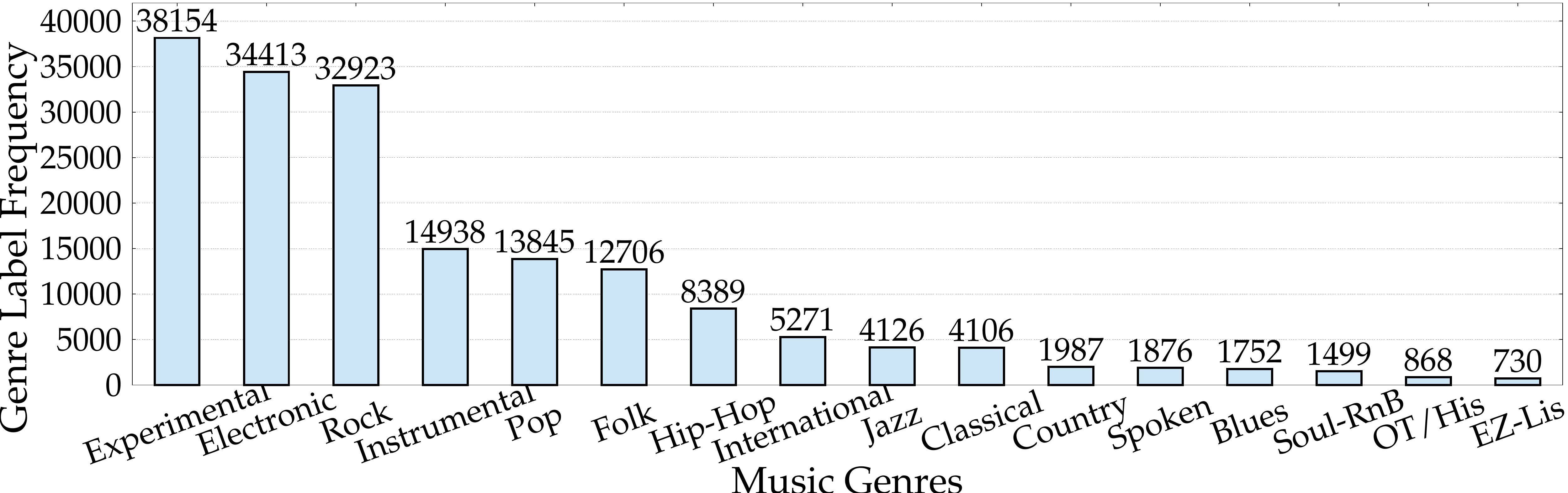}}
	\caption{Music Genre Distribution of FMA Dataset.}
	\label{fig:fma-long-tail}
\end{figure}

Long-tail music genres cannot be ignored as they contain rich musical information. Besides, the data with long-tail distribution is quit common in real-world settings. As a widely used music research dataset, Free Music Archive (FMA) can be used for evaluating the performance of MGC models. As demonstrated in Figure~\ref{fig:fma-long-tail}, nearly 82\% of the genres in FMA have only a few examples. It means that the MGC models need to be able to identify the genres with the limited number of training instances.

\begin{figure}[htbp]
	\centerline{\includegraphics[width=1.0\columnwidth]{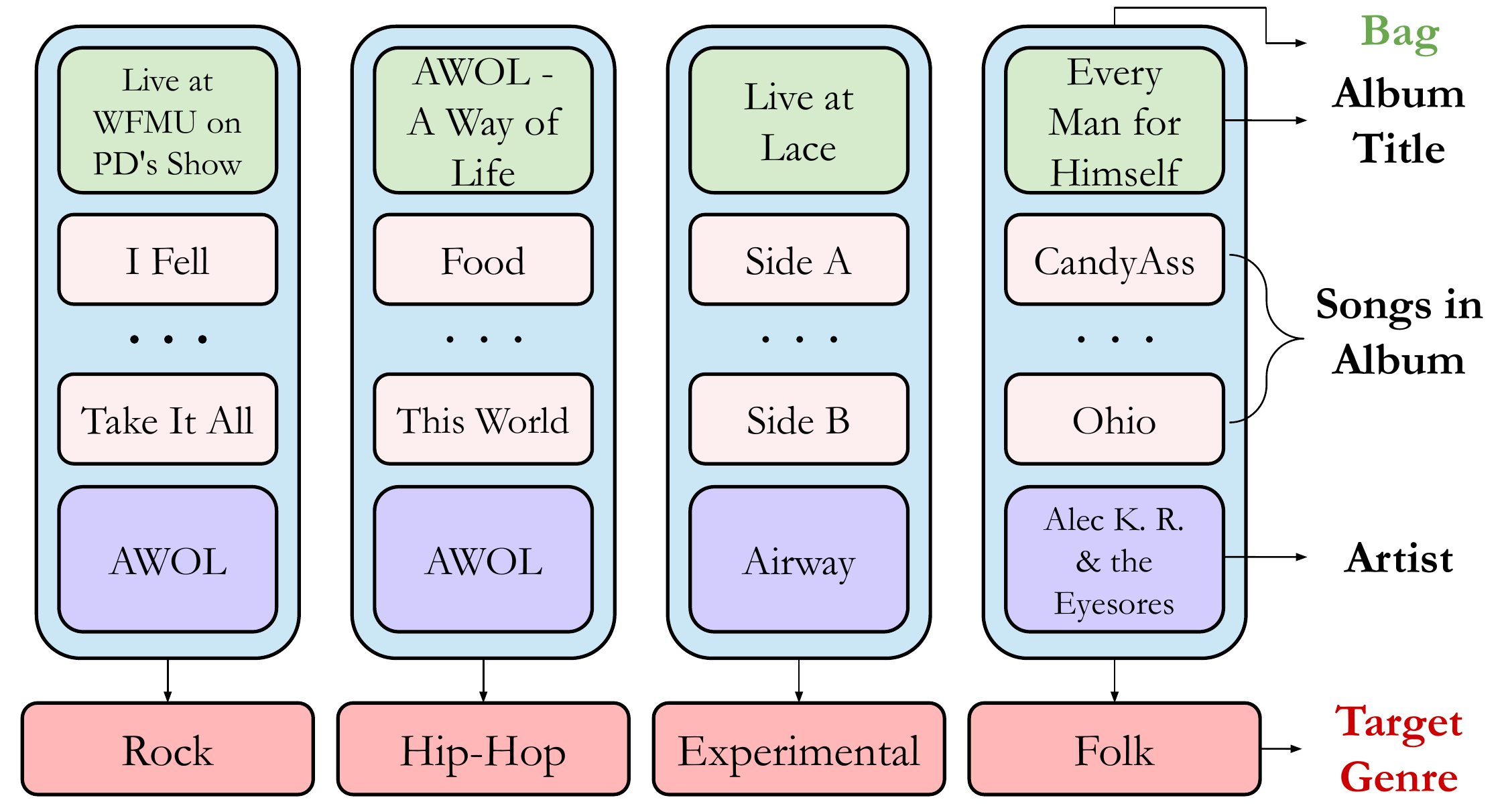}}
	\caption{An Example of MIL for Music Genre Classification.}
	\label{fig:mil-example}
\end{figure}

Inspired by the achievements of multi-instance learning and attention mechanism in various scenarios~\cite{zhou2004multi,DBLP:conf/dasfaa/LiuZGJ22}, we propose a novel multiple-instance attention mechanism (\tech) to accurately process those long-tail genres. Similar to the data preprocessing schema in multi-instance learning, we construct the corresponding bag-level dataset using the album-artist pair as the key of the music segment bag. Figure~\ref{fig:mil-example} shows the structure of our designed music segment bags for multi-instance learning. The music segments with the same album ID and artist ID would be put into the corresponding album-artist bag since the genres of the segments with the same album ID and artist ID would be the same. By leveraging the bag-level multiple-instance training mechanism, the long-tail problem should be alleviated. Besides, we employ a multiple-instance attention mechanism to help the neural networks identify the genres of the given music segments. With the multiple-instance attention mechanism utilized in various feature-based and neural network-based models, all of these models achieve better performance. {\em To the best of our knowledge, this is the first work to comprehensively evaluate the performance of various state-of-art music genre classification models on long-tail music genre classification benchmarks.}

The key contributions of this work are summarized as follows.

\begin{itemize}
    \item We propose the multiple-instance learning method for music genre classification, which alleviate the long-tail effect of the music genre distribution. The multiple-instance learning for music genre classification uses the album ID and artist ID pair as the key of the music bag, to build the bag-level dataset. By training models on the built training dataset, the models outperform those without adopting multiple-instance learning.
    \item We propose the \tech\ for the long-tail music genre. \tech\ can calculate the attention scores of each music segment in the bag to help identify the long-tail music genre of the given music segments. With the \tech, the neural networks-based models have better performance than other models without \tech\ adopt. 
    \item We conduct comprehensive experiments to evaluate the overall performance and long-tail music genre classification performance of the proposed \tech. We also use various metrics to evaluate data-imbalanced classification tasks to compare the performance of \tech\ with those of the state-of-art feature-based and neural networks-based methods. {\em The results demonstrate \tech\ achieves the state-of-the-art performance}. 
\end{itemize}

We arrange the rest of this paper as follows: In section II, we describe the related works about classifying balanced music genres and the long-tail genre classification. In section III, we present our proposed methods in detail. In Section IV, we describe the experiment design and analyze the results. The threads to validity are also included in this section. At last, we conclude this paper and discuss our future works.

\section{Related Works}

Conventional statistic learning-based models~\cite{DBLP:journals/ml/BergstraCEEK06,DBLP:conf/sigir/LiOL03,jiang2002music} are hard to meet the requirements of dealing with massive data in production environment. Recently, The deep learning-based music genre classification models~\cite{DBLP:conf/nips/LeePLN09,DBLP:conf/icassp/SigtiaD14,DBLP:conf/icassp/LeglaiveHB15,DBLP:conf/icassp/ChoiFSC17,DBLP:conf/aaaiss/GhosalS20,DBLP:conf/ijcnn/WangMFOL19,DBLP:journals/access/YangFWYL20,DBLP:journals/mta/LiuFLWL21} have been widely adopted for MCG and have achieved promising performances. In 2009, ~\cite{DBLP:conf/nips/LeePLN09} built a MGC model using convolutional deep belief network architectures. Then, ~\cite{DBLP:conf/icassp/SigtiaD14} leveraged the rectifier convolutional neural networks to extract informative features from audio data. ~\cite{DBLP:conf/icassp/LeglaiveHB15} used the bidirectional long short term memory (BiLSTM) and recurrent neural networks (RNN) on singing voice detection. The works, which were combined with different structures of neural networks, achieved better performance than the traditional neural networks. ~\cite{DBLP:conf/icassp/ChoiFSC17} leveraged both CNNs and RNNs in their works and introduce convolutional recurrent neural networks (CRNNs). The clustering augmented learning method (CALM), proposed by ~\cite{DBLP:conf/aaaiss/GhosalS20}, also achieved promising performance. The CALM is based on the concept of simultaneous heterogeneous clustering and classification to obtain more effective music representations from the audio features extracted via the LSTM autoencoder. ~\cite{DBLP:conf/ijcnn/WangMFOL19} proposed an improved technique called CRNN in Time and Frequency dimensions (CRNN-TF), which captures spatial dependencies of music signal in both time and frequency dimensions in multiple directions. Considering the aforementioned limitations, ~\cite{DBLP:journals/access/YangFWYL20} proposed a hybrid architecture, named the parallel recurrent convolutional neural network (PRCNN). ~\cite{DBLP:journals/mta/LiuFLWL21} proposed bottom-up broadcast neural network which transfers more suitable semantic features for the decision-making layer to discriminate the genre of the unknown music clip. These works mainly focus the class-balanced genre classification, regardless of the effect of long-tail genres.

However, to the best of our knowledge, there are only a few researches on long-tail MGC tasks~\cite{DBLP:conf/ismir/ChoiLPN19,DBLP:conf/iccbr/CrawHM15,DBLP:conf/flairs/ValerioPCBS18}. Choi et al~\cite{DBLP:conf/ismir/ChoiLPN19} proposed to leverage zero-shot learning to handle unseen labels such as newly added music genres or semantic words that users arbitrarily choose for music retrieval. Urbano et al~\cite{DBLP:conf/iccbr/CrawHM15}  exploited the combined knowledge, from audio and tagging, using a hybrid representation that extends the track’s tag-based representation by adding semantic knowledge extracted from the tags of similar music tracks. Valerio et al~\cite{DBLP:conf/flairs/ValerioPCBS18} proposed a resampling approach to face the class-imbalance problem applied to music genre classification. Although these works have made a positive exploration in the task of long-tail music genre classification, their performance is still not satisfactory. Here we adopt the multiple-instance attention mechanism for music to further improve the performance of MGC models.

\section{Methodology}

In this section, we first introduce the notations used in this work and briefly explain our workflow. Then we present our multiple-instance attention mechanism-based models for music genre classification step by step.

\begin{figure*}[h]  
	\centering
	\includegraphics[width=1.0\textwidth]{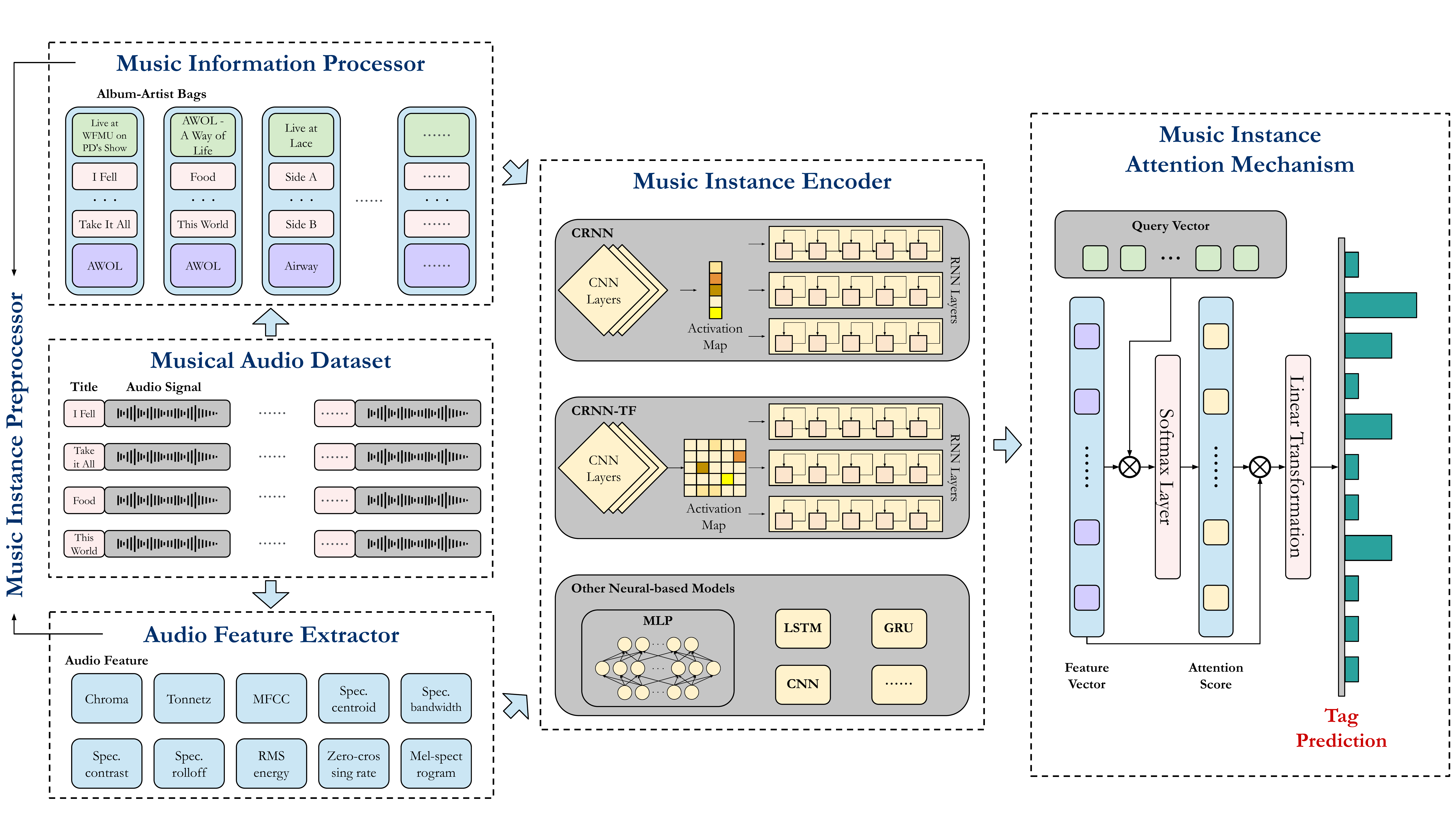}
	\caption{The Overall Workflow of the \tech-based Music Genre Classification Model.}
	\label{fig:workflow}
\end{figure*}

\subsection{Notations}      

Inspired by Multiple-instance Learning (MIL) from natural language processing field, we split all musical audio segments into multiple album-artist bags and denote them as $\{\mathcal{S}_{1},\mathcal{S}_{2},…\}$. Each bag $\mathcal{S_i}$ contains multiple instance $\{s_1, s_2, ...\}$ with the same artist ID $\mathcal{P}$ and album ID $\mathcal{A}$. Besides, each instance $s$ in these bags is encoded as mp3 format with a sample rate of 44.1KHz and a sample size of 320 kb/s with stereo channels.

\subsection{Overall Workflow}

Our model consists three major components as shown in Figure~\ref{fig:workflow}.

\paragraph{Music Instance Preprocessor} Given an instance and its related album ID and artist ID pair, we employ multi-feature extraction methods to embed music audio segments into continuous vector space. In this study, Log-amplitude Mel-spectrogram, MFCC, Chroma, and more than 10 audio feature extraction methods in total are used to extract music audio features.

\paragraph{Music Instance Encoder} The music audio encoder is responsible for encoding the low-dimensional music audio vector representation. The Convolutional Recurrent Neural Network is used to implement the music instance encoder for the neural network-based models. For the statistical learning-based model driven by music feature engineer, the music instance is omitted because the low-dimensional features are not suitable for training.

\paragraph{Multi-instance Attention Mechanism} Under the guidance of the final audio embeddings, \tech~ can identify the most informative music segment exactly matching the relevant genre. 

\subsection{Music Instance Preprocessor}

\paragraph{Music Information Processor}
The music information processor aims to process the music dataset and converts them to the corresponding bag level. This process follows the principle of MIL. Specifically, the processor splits all musical audio segments into multiple album-artist bags, where each music segment in the bag shares the same album ID and artist ID. The reason why we leverage MIL to process the music dataset is that the music segments with the same album ID and artist ID always belong to the same genre, which is observed at whole FMA dataset.

\paragraph{Audio Feature Extractor}

The audio feature extractor is designed to embed music segments into continuous digital representations (i.e. vectors). 
We extract features from the music audio by 10 methods, which are Chroma, Tonnetz, MFCC, Spec. centroid, Spec. bandwidth, Spec. contrast, Spec. rolloff, RMS energy, Zero-crossing rate feature, and log-amplitude mel-spectrogram. 
We implement 9 of above methods via Python library -- Librosa.
As for log-amplitude mel-spectrogram, its output is a $96\times 1360$ matrix of mel-spectrogram, where each row and column of this matrix corresponds to a Mel-frequency scale and a Mel-frequency time frame, respectively.

\subsection{Music Instance Encoder} 

The encoding layer aims to convert given instances into their latent vectors and maintains their semantics at the same time.
In this work, we choose neural networks with convolutional layers, e.g. the CRNN and CRNN-TF, to encode input embeddings extracted from the log-amplitude mel-spectrogram feature extraction method. 
The reason is that the mel-spectrogram feature has a higher dimension compared with other types of features, which can be smoothly computed by CRNN and CRNN-TF.
For other features obtained from the other 9 feature extraction methods, due to the small size of dimensions, they can be processed by both CRNN/CRNN-TF and other encoding methods such as MLP.
Other networks such as recurrent neural networks~\cite{DBLP:conf/icassp/LeglaiveHB15} can also serve as audio encoders.

\paragraph{CRNN} The convolutional recurrent neural network (CRNN) consists of convolutional layers and Gated recurrent unit (GRU) layers. GRU is a gating mechanism in RNN. The GRU and RNN have very similar structures. But, GRU has fewer parameters by dropping the traditional output gate.  

\paragraph{CRNN-TF} The convolutional recurrent neural network in Time and Frequency dimensions (CRNN-TF) is a variant of CRNN. It can extract spatial dependencies in both Time and Frequency dimensions of music signals. CRNN-TF has achieved promising performances in several state-of-the-art deep learning-based music models.

\subsection{Multi-instance Attention Mechanism} 

Given the musical segment embeddings $s_{p, a} = \{$$s_{1}$, $s_{2}$, ..., $s_{m}$$\}$, we apply a plain selective attention over them to get the musical genre representation $q_{g}$ for classifying the genres. We adopt $q_{g}$ as attention query vector initialed by xavier uniform. The attention for each musical segment in $s_k$ is defined as follow:
\begin{equation}
	e_{k}= {\rm tanh}(W_s [s_k; q_{g}]) + b_s 
\end{equation}
\begin{equation}
	a_{k} = \frac{\exp{(e_{k}})}{\sum_{j=1}^{m} \exp{(e_{j}})}
\end{equation}
where $[x_1; x_2]$ denotes the vertical concatenation of $x_1$ and $x_2$, $W_s$ is the weight matrix, and $b_s$ is the bias. The converged nodes will share the parameters. By doing so, we can compute attention operations on each label of the music segments to obtain their corresponding musical genre representations.
\begin{equation}
	g_{p, a} = {\rm ATT}(q_{g}, {s_1, s_2, ..., s_m})
\end{equation}
The musical genre representation $g_s$ will be fed to compute the conditional probability $P(g|h, t, s)$, which is shown as below:
\begin{equation}
	P(g|h, t, S_{p, a})=\frac{\exp(o_r)}{\sum_{\hat{g}\in\mathcal{G}}\exp{(o_{\hat{g}})}}\label{eq}
\end{equation}
\begin{equation}
	o = M g_{p, a}
\label{eq:final-one}
\end{equation}
where $o$ is the scores of the music genres. We then use a discriminative matrix $M$ to obtain theses genre scores, as Equation~\ref{eq:final-one} shown.

\section{Experiments}

In this section, we first introduce the experimental settings. Second, we evaluate our proposed methods and various baseline models in terms of overall and long-tail classification performance. To comprehensively understand how MATT affects the performance of models in identifying the music segment without valid album ID or artist ID, we conduct a case study to explain it. In the end, we introduce the threats that may affect the validity. 

\subsection{Experimental Setting}

\subsubsection{Datasets} 

We evaluate our proposed methods on the FMA dataset which is arranged in a taxonomy of 16 genres. The FMA dataset is a representative dataset used to evaluate the performance of the MGC model~\cite{DBLP:conf/icassp/ChoiFSC17,DBLP:conf/ijcnn/WangMFOL19}. Depending on the number of samples in the dataset, the FMA dataset is available for 4 datasets with different scales, such as full, large, medium, and small datasets. We utilize the 16 classes medium dataset, which demonstrates long-tail distribution, for evaluation.


\subsubsection{Metrics and Evaluation Procedure}

For evaluation, we draw precision-recall curves for all models. To further verify the performance of our model for long-tail genres, we report the Top@K Accuracy results. The baseline dataset and codes can be found in Github\footnote{\ \url{https://github.com/mdeff/fma}} \footnote{\ \url{https://github.com/FishInMedi/CRNN-TF}}.

\begin{figure*}[h]
\centering
\subfigure[Precision-recall curves of the proposed methods vs feature-based baselines.]
{
    \includegraphics[width=0.32\textwidth]{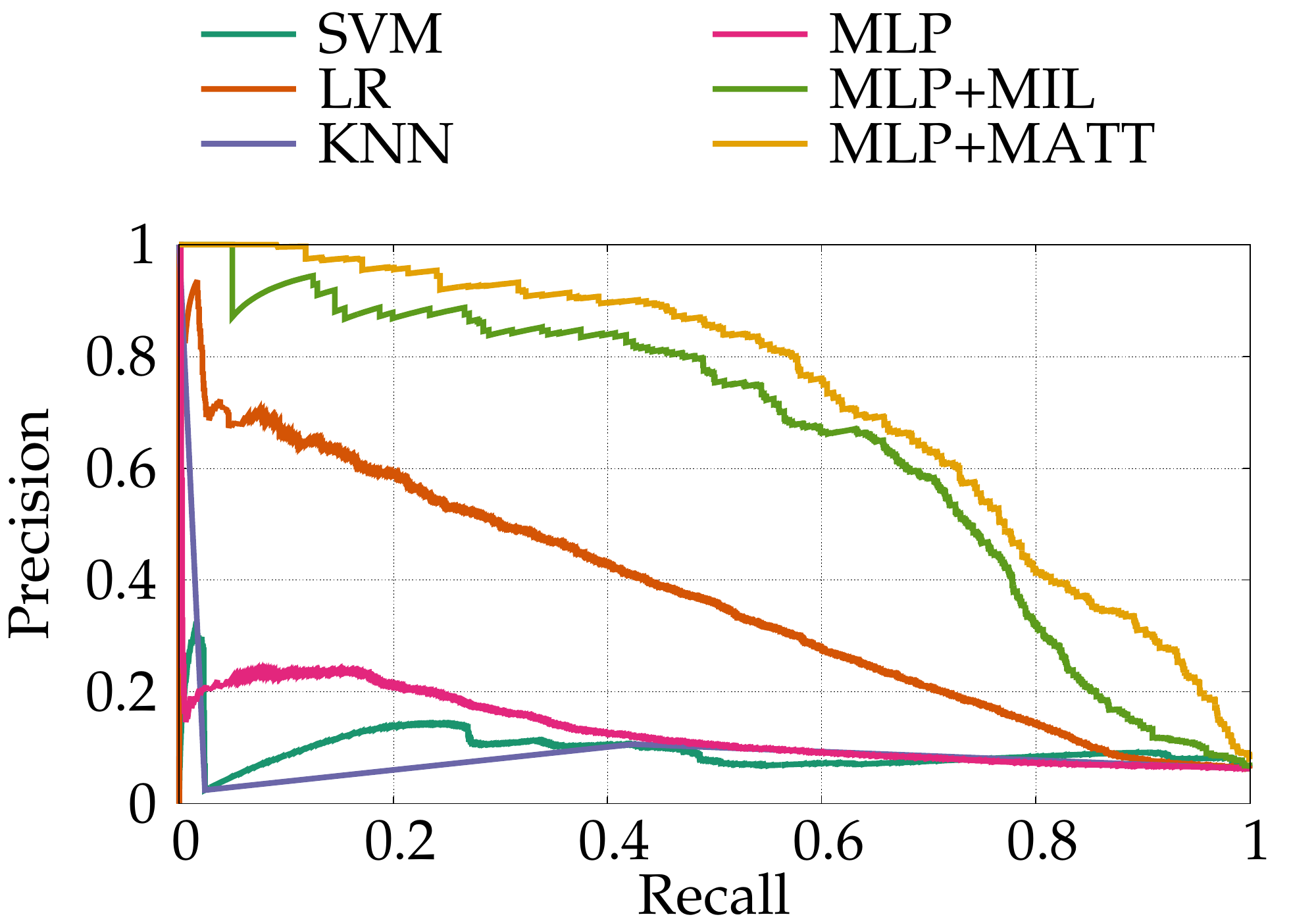}
    \label{fig:pr1}
}
\subfigure[Precision-recall curves of the proposed methods vs neural-based baselines.]
{
    \includegraphics[width=0.32\textwidth]{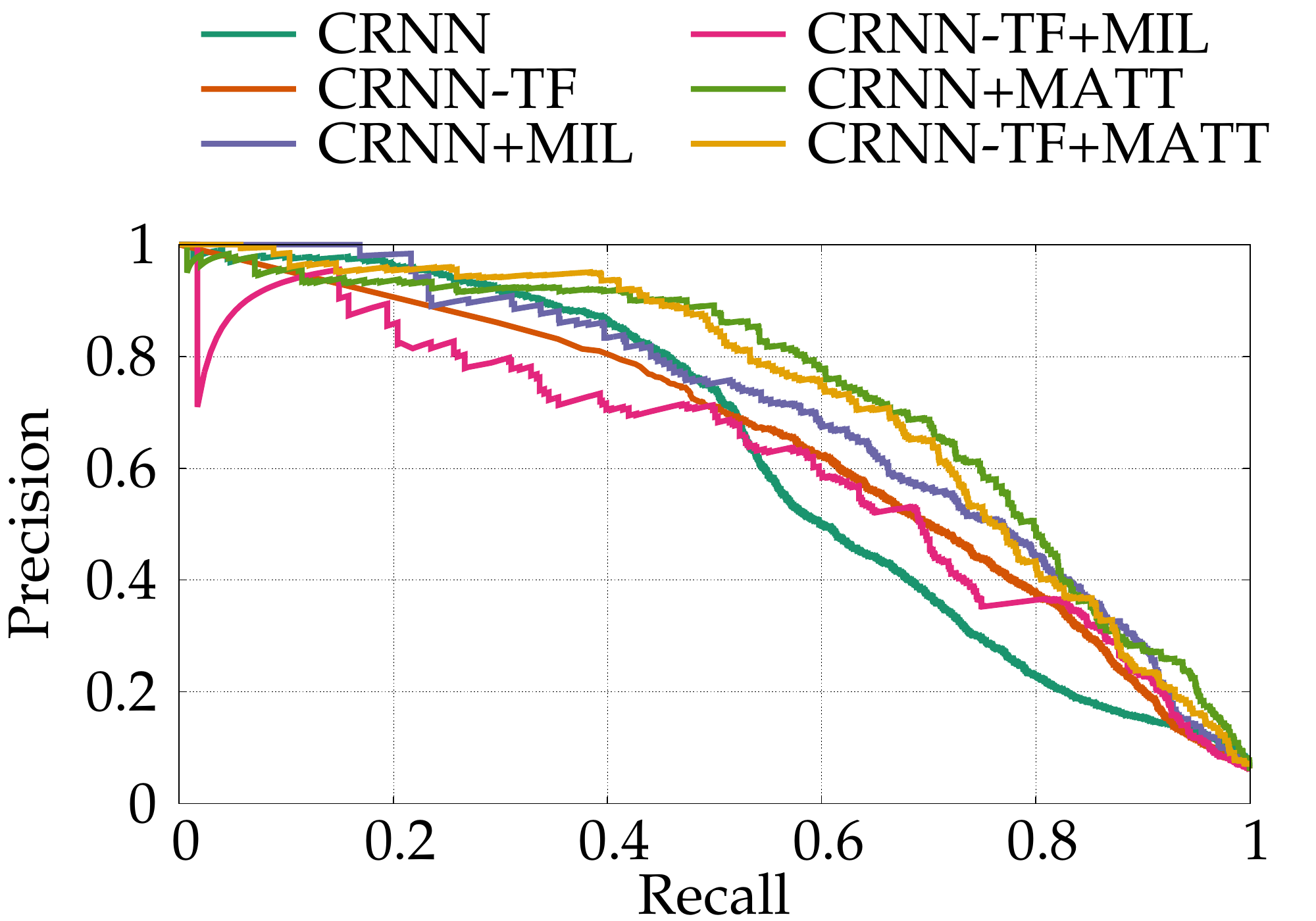}
    \label{fig:pr2}
}
\subfigure[Testing performance of neural-based models on the FMA Dataset.]
{
    \includegraphics[width=0.25\textwidth]{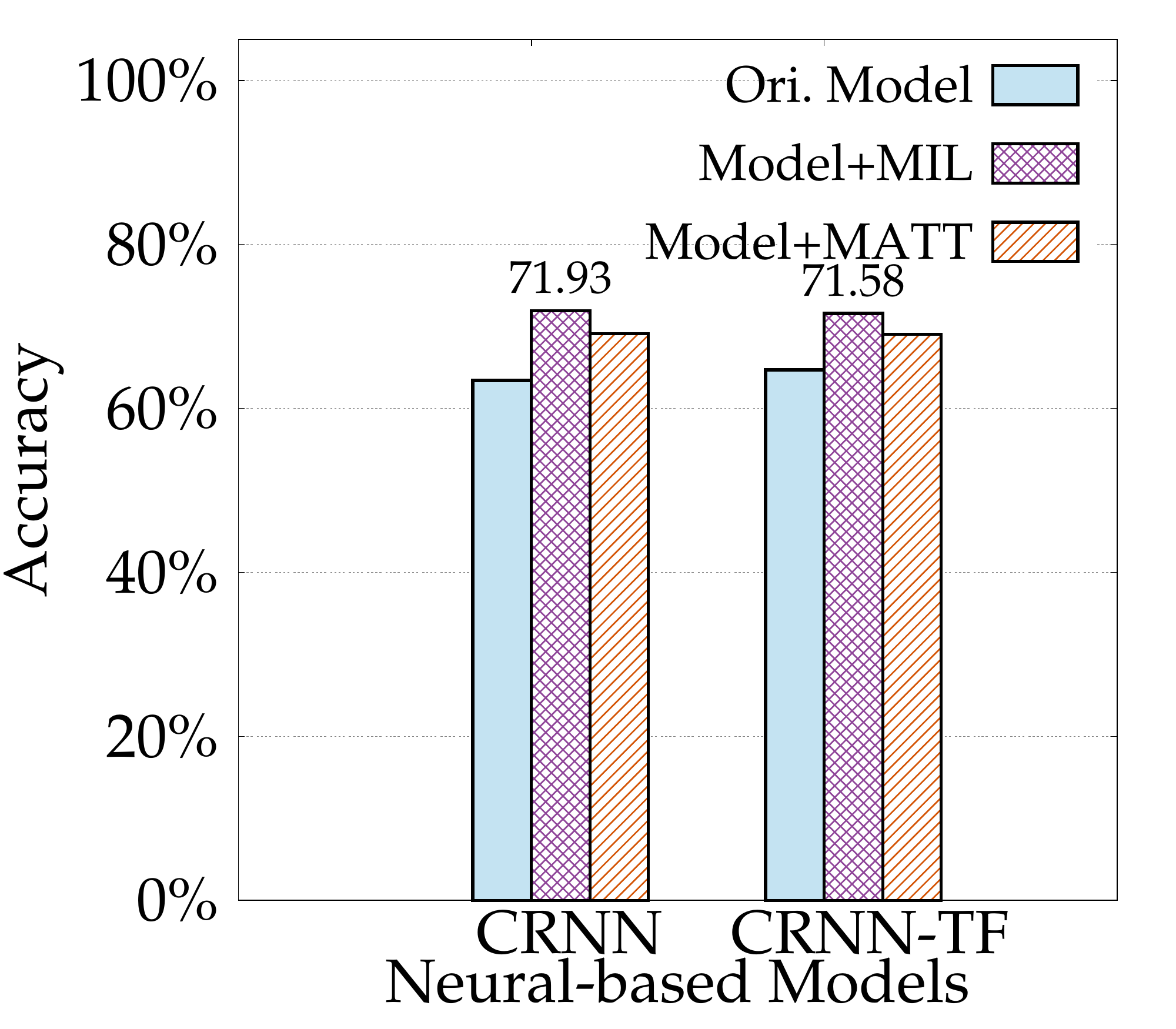}
    \label{fig:accuracy2}
}
\caption{PR-Curve and Accuracy of Our Proposed Models against Baselines}
\end{figure*}




\subsubsection{Comparison Models}

For baselines, we evaluate both neural network-based and feature-based models.
To verify the performance of multiple-instance learning mechanism in music genre classification, we report the results of our method with various baseline models including \lr, \svm, \knn\, \mlp, \textbf{CRNN}, \textbf{CRNN-TF}, and the same model equipped with various advanced learning strategies. 
Among these strategies, \textbf{+MIL} is the multi-instance learning strategy over musical audio segments and \tech\ is our proposed multi-instance attention mechanism for music genre classification. 
For the models, \lr, \svm, \knn\, \mlp\ use the 1-9 feature sets in table 1 as the input, while \crnn\ and \cgrnn\ adopt log-amplitude mel-spectrogram features as the input.

\subsubsection{Hyperparameter Setting and Reproducibility}
To make the results reproducible, we adopt the default data split schema proposed by FMA and the default hyperparameter settings from baseline methods. 

\subsection{Overall Evaluation Results}


To evaluate the performance of our proposed model, we compare the precision-recall curves and accuracy of our model with various baseline models. The accuracy results are demonstrated in Table 1 and Figure 6. As shown in both Table 1 and Figure~\ref{fig:pr1} and~\ref{fig:pr2}, we observe that:
(1) For the feature engineering-based models, the accuracy of MLP models with MIL and MATT mechanisms outperforms the other baseline models. We also observe similar trends in CRNN and CRNN-TF. These results confirm that the MIL and MATT can significantly improve the model performance.
(2) The accuracy of models with MATT is slightly lower than those of models with MIL mechanisms. Notice that the distribution of the dataset is long-tail, so only the accuracy metric cannot help conclude the performance of the models. 

To effectively evaluate the long-tail genres, we adopt the precision-recall curve, which is one of the most widely-adopted evaluation metrics in information retrieval field, to help evaluate the models better.The precision-recall curves are shown in Figure~\ref{fig:pr1} and~\ref{fig:pr2}. We can conclude from both figures: 
\begin{table}[htbp]
\caption{Testing Accuracy (\%) of various features and models on the FMA Data Medium Subset}
\begin{center}
\scalebox{0.78}{
\begin{tabular}{lccccccc}
\toprule
\textbf{Feature Set}& \textbf{Dim.} & \textbf{LR} & \textbf{KNN} & \textbf{SVM} & \textbf{MLP} & \textbf{MLP+MIL} & \textbf{MLP+MATT} \\
\midrule
1 Chroma & 84 & 44 & 44 & 48 & 49 & \textbf{47.74} & 39.53 \\

2 Tonnetz  & 42 & 40 & 37 & 42 & 41 & \textbf{45.90} & 43.45\\

3 MFCC & 140 & 58 & 55 & 61 & 53 & \textbf{64.71} & 63.19\\

4 Spec. Centroid & 7 & 42 & 45 & 46 & 48 & \textbf{49.36} & 44.93\\

5 Spec. BW.. & 7  & 41 & 45 & 44 & 45 & 43.26 &\textbf{ 44.23} \\

6 Spec. Contrast & 49  & 51 & 50 & 54 & 53 &\textbf{55.69} & 49.59 \\

7 Spec. Rolloff & 7  & 42 & 46 & 48 & 48 & 49.16 & \textbf{49.20} \\

8 RMS Energy & 7 & 37 & 39 & 39 & 39 &\textbf{41.08} & 38.67 \\

9 Zero-Crossing & 7 & 42 & 45 & 45 & 46 & \textbf{47.96} &  46.91\\

3+6& 189 & 60 & 55 & 63  & 54 & 65.68 & \textbf{68.91}\\

3+6+4& 196 & 60 & 55 & 63  & 53 & \textbf{65.84} & 63.31\\

1 to 9& 518 & 61 & 52 & 63  & 58 & \textbf{69.22} & 68.40 \\
\bottomrule
\end{tabular}
\label{AccuracyTest}
}
\end{center}
\end{table}
(1) All models equipped with MATT mechanism can achieve good precision while their recalls are smaller than 0.6. 
The feature engineering-based models without equipping any advanced mechanisms show the worst performance among all the models. Besides, the models equipped with MATT mechanism outperform those equipped with MIL. The MATT-equipped MLP and CRNN-TF achieve the best performance under different feature extraction methods. In addition, CRNN-TF achieves the best performance among all the models.
(2) The models equipped with MIL mechanism are better than their corresponding baseline models, except for the CRNN-TF. Such phenomena reveal that although the CRNN-TF with MIL achieves state-of-the-art accuracy, the performance of CRNN-TF in music information retrieval benchmarks with long-tail distribution should still be questioned. However, the attention-based CRNN-TF with MIL helps address this issue.

From these experimental results, we can conclude that the MATT-based models achieve the best performances compared with other models. Considering the accuracy, precision, and recall at the same time, CRNN-TF with MATT achieves the best performance.

\subsection{Evaluation Results for Long-tail Genres}

\begin{table}[htbp]
	\centering
\caption{Top@K Accuracy (\%) on long-tail classes.}
	\scalebox{0.8}{
		\begin{tabular}{lcccccc}
			\toprule
			 \textbf{Number of}  &
			 \multicolumn{3}{c}{\multirow{2}{*}{\textbf{$<$100}}} & \multicolumn{3}{c}{\multirow{2}{*}{\textbf{$<$200}}} \\
 		     \textbf{Training Instances}  &
			 \multicolumn{3}{c}{} & 
			 \multicolumn{3}{c}{} \\
	        \midrule
			 \textbf{Top@K} & \textbf{K=2} & \textbf{K=3} & \textbf{K=5} & \textbf{K=2} & \textbf{K=3} & \textbf{K=5} \\
			\midrule
			1 LR     	& $<$5.0  & 7.06 & 25.29 & $<$5.0 & 6.38& 22.87   \\
			2 KNN    & 17.06  & 25.88 & 45.29  & 15.43 & 23.94 & 41.49  \\
			3 SVM     	& 11.76  & 19.41 & 44.71 & 10.64 & 17.55 & 40.42   \\
			\midrule
			4 MLP     	& $<$5.0  & 10.59 & 28.23 & $<$5.0 & 10.11 & 27.13   \\
    		5 MLP+MATT   &  \textbf{28.24}  &  \textbf{44.71} & \textbf{47.06} & \textbf{26.60} & \textbf{41.49} &  \textbf{44.15}   \\
			\midrule
			6 CRNN     	& 9.41  & 13.53 & 34.71 & 8.51 & 12.23 & 38.83   \\
    		7 CRNN + MATT   &  \textbf{54.12}  &  \textbf{55.88} & \textbf{57.65} & \textbf{49.47} & \textbf{51.60} &  \textbf{61.7}   \\
			\midrule
			8 CRNN-TF     	&$<$5.0   &  9.41 & 20.59  & $<$5.0 & 8.51 & 18.62   \\
    		9 CRNN-TF + MATT   &  \textbf{54.12}  &  \textbf{63.53} & \textbf{65.29} & \textbf{59.04} & \textbf{61.17} &  \textbf{71.21}   \\
			\bottomrule
		\end{tabular}
	}	
\end{table}

To further demonstrate the improvements in performance for long-tail genres, we extract a subset of the testing dataset in which all the genres have fewer than 100/200 training instances. 
We employ the Top@K metric for evaluating the results. 
For each album-artist pair, the evaluation requires its corresponding golden genre in the first K candidate relations recommended by the models. 
Since it is difficult for any of the existing models to extract long-tail genres, we select K from {2, 3, 5}. 
We report the micro average Top@K accuracy for these subsets in Table 2. Similar to former contexts, the 1-5 models use the 1-9 feature set in Table 2 as the input, while the 6-9 models adopt log-amplitude mel-spectrogram features as the input. From this table, we have the following observations: 
(1) Tough the models without \tech\ achieves high accuracy, their performance in predicting the genres of long-tail music degrades dramatically. It reveals that conventional works, both feature engineering-based and neural-based works, are suffering from an extreme long-tail problem.
(2) For both feature engineering-based Models and neural-based models, the models with our \tech\  outperform other baseline models in extracting long-tail music genres. Among the \tech\ based works, the CRNN-TF with \tech\ achieves the best performance. 
(3) The neural-based models outperform the feature engineering-based models in identifying data-poor genres. It confirms that the neural-based models, which automatically extract music features from the Log-amplitude Mel-spectrogram data, have more potential in long-tail music genre identification than the complex feature engineering-based models.
(4) Although our MATT mechanism has achieved better results in the task of long-tail music genre classification as compared with other SOTA methods, the results of all the ML and DL algorithms with MATT are still not satisfying. In the near future, we plan to adopt more advanced schemes and introduce extra information to solve this problem.

\subsection{Case Study}

We noticed that the bag-level evaluation requires extra information, such as album ID and artist ID. 
However, in some scenarios, this information is hard to obtain. 
Thus we carry out a case study on segment-level evaluation. 
The segment-level evaluation does not need any extra information but the music segment in the evaluation process. 
In the training phase, since the album ID and artist ID is easy to obtain, we train the models using the processed bag-level dataset. 
In the testing phase, we evaluate the models without using the multi-instance learning strategy. 
In other words, the model is only equipped with a plain attention mechanism in the evaluation stage.

We conduct experiments using MLP with the whole FMA feature sets.
By training the MLP model on the bag-level dataset using features 1 to 9 from the feature set, the testing accuracy degrades from 68.83\% to 63.69\%. Nevertheless, we argue that this degradation in testing accuracy is acceptable. 
Because in the testing phase, no album ID or artist ID is used to help identify the genre information. 
In addition, we found that the testing accuracy is still higher than that of the training model at the segment level, whose testing accuracy is only 58\%. 
For the rest features from the feature set, we observed a similar trend in terms of accuracy and average precision. 

\subsection{Threats to Validity}
\begin{enumerate}
\item ML Platform: We implement MATT on an industrial-grade ML platform, Pytorch. We notice the baseline works were implemented in Scikit-learn and Keras. For the metrics not reported in the baseline works,  we extract the parameters of the models and try our best to keep the parameters the same as before. For the metrics has been reported in previous works,  we adopt the results from baseline works to avoid the possible evaluation bias.  
\item Hardware: We notice that different hardware may cause differences in the performance of music genre classification models, so we declare the hardware we use in our experiments. In our study, we use a Linux server with two 48-core Intel CPUs and 376 GB of memory. The machine is also equipped with four NVIDIA RTX 3090 GPUs. 

\end{enumerate}

\section{Conclusion}

In this paper, we propose the MATT mechanism to identify music genres, especially the long-tail genres, in a large-scale music benchmark. 
Comprehensive experimental results demonstrate that the MATT can significantly improve the performance of the MGC models for identifying long-tail music genres. 
Moreover, MATT can enhance the models' capability of classifying music genres in both feature engineering- and neural network-based music genre classification models. 
Evaluation results on the segment-level testing dataset also demonstrate that the MATT is still competitive even when album ID and Artist ID are missed.

\bibliographystyle{IEEEtran}
\bibliography{references} 
\end{document}